\documentclass[aps,twocolumn,preprintnumbers,amsmath,amssymb,array]{revtex4}

\usepackage{graphicx}% Include figure files
\usepackage{dcolumn}% Align table columns on decimal point
\usepackage{bm}% bold math

\DeclareGraphicsExtensions{.ps,.eps}

\let\footnote=\endnote

\begin{document}
\title{Inelastic effects in Aharonov-Bohm molecular interferometer}

\author{$\mbox{Oded Hod}^{1\dagger}$, $\mbox{Roi Baer}^2$, and
$\mbox{Eran Rabani}^1$}

\affiliation{$^1$School of Chemistry, Tel Aviv University, Tel Aviv
69978, Israel; $^2$Institute of Chemistry, The Hebrew University of
Jerusalem, Jerusalem 91904 Israel.}

\date{\today}

\begin{abstract}
Inelastic effects arising from electron-phonon coupling in molecular
Aharonov-Bohm (AB) interferometers are studied using the
nonequilibrium Green's function method.  Results for the
magnetoconductance are compared for different values of the
electron-phonon coupling strength.  At low bias voltages, the coupling
to the phonons does not change the lifetime and leads mainly to
scattering phase shifts of the conducting electrons.  Surprisingly,
opposite to the behavior of an electrical gate, the magnetoconductance
of the molecular AB interferometer becomes more sensitive to the
threading magnetic flux as the electron-phonon coupling is increased.
\end{abstract}

\maketitle

Control of conductance in molecular junctions is of key importance in
molecular electronics~\cite{Joachim2000,Nitzan2003_1}.  The current in
these junctions is often controlled by an electrical gate designed to
shift conductance peaks into the low-bias regime.  Magnetic fields on
the other hand, have been rarely used due to the small magnetic flux
captured by molecular conductors (an exception is the Kondo effect in
single-molecule transistors~\cite{Park2002,Liang02}). This is in
contrast to a related field, electronic transport through mesoscopic
devices, where considerable activity with magnetic fields has led to
the discovery of the quantum hall effect~\cite{Klitzing80} and a rich
description of transport in such
conductors~\cite{Webb1985,Timp1987,Yacoby1995,Oudenaarden1998,Fuhrer2001}.
The scarcity of experimental activity is due to the belief that
significant magnetic response is obtained only when the magnetic flux,
$\phi$, is on the order of the quantum flux, $\phi_0=h/e$ (where $e$
is the electron charge and $h$ is Planck's constant). Attaining such a
flux for molecular and nanoscale devices requires unrealistically huge
magnetic fields~\cite{Hod2004}.

Recently, we have described the essential physical requirements
necessary for the construction of nanometer scale magnetoresistance
devices based on an AB~\cite{Aharonov1959} molecular
interferometer~\cite{Hod2004,Hod2005}. The basic idea was to weakly
couple a molecular ring to conducting leads, creating a resonance
tunneling junction.  The resonant state was tuned by a gate potential
to attain maximal conductance in the absence of a magnetic field. The
application of a relatively small magnetic field shifts the state out
of resonance, and conductance was strongly suppressed within fractions
of the quantum flux.  The combination of a gate potential and a
magnetic field reveals new features and provides additional
conductivity control~\cite{Hod2005_1,Hod2006}.

Our previous study has neglected completely inelastic effects arising
from electron-phonon
interactions~\cite{Ness1999,Segal2002,Sutton2002,Hanggi2003,Ueba2003,Galperin2004,Mitra2004}.
Can a relatively small magnetic flux change significantly the
conduction in molecular rings when the electron-phonon coupling
becomes significant?  Or, perhaps inelastic effects will broaden the
resonant state and conduction will not be suppressed significantly
upon the application of the magnetic field.  The decay of the
amplitude of the AB oscillations due to electron-phonon coupling has
been studied for mesoscopic systems~\cite{Guinea2002}. In this letter
we address this problem for molecular rings, where we focus on the low
range of the magnetic flux appropriate for molecular rings. Opposite
to the effects of an electrical gate, we find that inelastic effect
arising from electron-phonon couplings narrows the magnetoconductance
peaks.

\begin{figure}
\begin{center}
\includegraphics[width=7cm]{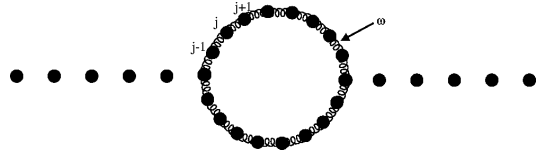}
\end{center}
\caption{A sketch of the AB ring. Each site on the ring can be
occupied by a single electron.  The ring sites are connect by springs
with a frequency $\Omega$.  An electron on site $j$ is coupled to the
local motion of this site, with a coupling strength $M$.}
\label{fig:sketch}
\end{figure}

We consider a two terminal junction of an Aharonov-Bohm ring with $N$
sites as sketched in Fig.~\ref{fig:sketch}.  A realization of this
model to realistic molecular loops is described
elsewhere~\cite{Hod2004,Hod2005,Hod2005_1,Hod2006}. We describe the
electronic structure of the ring and the leads using a magnetic
extended H{\"u}ckel model~\cite{Hod2004,Hodthesis}. The description of
the ring also includes local phonons and electron-phonon interactions
are approximated to lowest order.  The phonon local frequency $\Omega$
and the coupling $M$ of an electron on site $j$ to the local motion of
site $j$ are the only two free parameters of the model.  The full
Hamiltonian in second quantization is given by:
\begin{equation}
\begin{split}
&H = \sum_{i,j} t_{i,j}(B) c_{i}^{\dagger} c_{j} + \sum_{m,n \in L,R}
\epsilon_{m,n}(B) d_{m}^{\dagger} d_{n} +\\ &\left(\sum_{m,j}
V_{m,j}(B) d_m^\dagger c_j + H.c.\right) + \sum_{k=0}^{N-1} \hbar
\omega_{k} (b_{k}^{\dagger} b_{k}+1/2) \\ &+ \sum_{j,k=0}^{N-1}
M_{j}^{k} c_{j}^{\dagger} c_{j} (b_{k}^{\dagger} + b_{k}).
\end{split}
\label{eq:hamil}
\end{equation}
The first two terms on the right hand side (RHS) of Eq.~\ref{eq:hamil}
represent the zero-order electronic Hamiltonian of the ring and leads,
respectively. $c_j^\dagger$ ($c_j$) is Fermion creation (annihilation)
operators of an electron on site $j$ on the ring, and $d_m^\dagger$
($d_m$) is Fermion creation (annihilation) operators of an electron on
site $m$ on the left (L) or right (R) lead.  $t_{i,j}(B)$ and
$\epsilon_{i,j}(B)$ are the hopping matrix elements between site $i$
and site $j$ on the ring and lead, respectively.  The third term on
the RHS of Eq.~\ref{eq:hamil} corresponds to the coupling between the
ring and the leads, where $V_{m,j}(B)$ is the hopping element between
site $m$ on the lead and site $j$ on the ring.  All hopping elements
depend on the magnetic field $B$, which is taken to be uniform in the
direction perpendicular to the ring plane.  Both linear and quadratic
terms in the magnetic field are included in the
calculation~\cite{Hod2004}. The last two terms in Eq.~\ref{eq:hamil}
represents the Hamiltonian of the phonons and the electron-phonon
interactions.  $b_{k}^{\dagger}$ ($b_{k}$) is a boson creation
(annihilation) operator of phonon mode $k$ with a corresponding
frequency $\omega_{k}$.  This set of phonon modes was obtained by a
unitary transformation from local to normal coordinates of a one
dimensional chain of coupled harmonic oscillators, characterized by a
single frequency $\Omega$, as illustrate in Fig.~\ref{fig:sketch}.
These frequencies constitute a band of width proportional to the
coupling between the oscillators.  The electron-phonon coupling is
approximated to lowest order.  Each site on the ring is coupled to all
phonon modes with a coupling strength $M_{j}^{k}=M
\sqrt{\frac{\Omega}{\omega_{k}}} U_{jk}$, where $U_{jk}$ are the
matrix elements of transformation matrix ${\bf U}$ from local to
normal modes.

The calculation of the conductance is described within the framework
of the nonequilibrium Green's function (NEGF)
method~\cite{Datta_book}. The total current $I=I_{el}+I_{inel}$ is
recast as a sum of elastic ($I_{el}$) and inelastic ($I_{inel}$)
contributions given by~\cite{Galperin2004,Mitra2004,Paulsson2005}
\begin{equation}
\begin{split}
I_{el}=\frac{2e}{\hbar}&\int\frac{d\epsilon}{2\pi}
\left[f(\epsilon,\mu_{\mbox \tiny R})-f(\epsilon,\mu_{\mbox \tiny
L})\right]\\& \mbox{Tr}\left[{\bf \Gamma}_{\mbox \tiny
L}(\epsilon){\bf G}^{r}(\epsilon) {\bf \Gamma}_{\mbox \tiny
R}(\epsilon){\bf G}^{a}(\epsilon)\right]
\end{split}
\label{eq:elastic}
\end{equation}
and
\begin{equation}
\begin{split}
I_{inel}=\frac{2e}{\hbar}&\int\frac{d\epsilon}{2\pi}\mbox{Tr}\left[{\bf
\Sigma}_{\mbox \tiny L}^<(\epsilon) {\bf G}^{r}(\epsilon) {\bf
\Sigma}_{ph}^{>}(\epsilon) {\bf G}^{a}(\epsilon)\right.\\&\left.
-{\bf \Sigma}_{\mbox \tiny L}^>(\epsilon) {\bf G}^{r}(\epsilon) {\bf
\Sigma}_{ph}^<(\epsilon) {\bf G}^{a}(\epsilon)\right],
\label{eq:inelastic}
\end{split}
\end{equation}
respectively.  The retarded (advanced) GFs satisfy the Dyson equation
\begin{equation}
{\bf G}^{r,a}(\epsilon) = \left\{[{\bf g}^{r,a}(\epsilon)]^{-1} - {\bf
\Sigma}_{\mbox \tiny L}^{r,a}(\epsilon) - {\bf \Sigma}_{\mbox \tiny
R}^{r,a}(\epsilon) - {\bf \Sigma}_{ph}^{r,a}(\epsilon)\right\}^{-1},
\label{eq:dyson}
\end{equation}
where ${\bf g}^{r,a}(\epsilon)$ is the uncoupled retarded (advanced)
electronic GF of the ring.  The greater (lesser) GFs satisfy the
Keldysh equation at steady state (for an initial noninteracting state)
\begin{equation}
{\bf G}^{\lessgtr}(\epsilon) = {\bf G}^{r}(\epsilon) \left[{\bf
\Sigma}_{\mbox \tiny L}^{\lessgtr}(\epsilon) + {\bf \Sigma}_{\mbox \tiny
R}^{\lessgtr}(\epsilon) + {\bf \Sigma}_{ph}^{\lessgtr}(\epsilon)\right] {\bf
G}^{a}(\epsilon).
\label{eq:keldysh}
\end{equation}
In the above equations, ${\bf \Sigma}_{\mbox \tiny
L}^{r,a,\lessgtr}(\epsilon)$, ${\bf \Sigma}_{\mbox \tiny
R}^{r,a,\lessgtr}(\epsilon)$, and ${\bf
\Sigma}_{ph}^{r,a,\lessgtr}(\epsilon)$ are the retarded ($r$),
advanced ($a$), lesser ($<$) and greater ($>$) self-energies arising
from the coupling to the left lead, right lead, and the phonons,
respectively, and ${\bf \Gamma}_{\mbox{\tiny
L,R}}(\epsilon)=i\left[{\bf \Sigma}_{\mbox{\tiny
L,R}}^{r}(\epsilon)-{\bf \Sigma}_{\mbox{\tiny
L,R}}^{a}(\epsilon)\right]$, where
\begin{equation}
{\bf \Sigma}_{\mbox{\tiny L,R}}^{r,a}(\epsilon)= \left(\epsilon {\bf
S}^{*}- {\bf V}^{*}(B)\right) {\bf g}_{\mbox{\tiny
L,R}}^{r,a}(\epsilon) \left(\epsilon {\bf S} - {\bf V}(B)\right).
\label{eq:SigmaLRra}
\end{equation}
In the above, ${\bf V}(B)$ is the lead-ring hopping matrix with
elements $V_{mj}(B)$, ${\bf S}$ is the overlap matrix between the
states on the leads and on the ring, and ${\bf g}_{\mbox{\tiny
L,R}}^{r,a}(\epsilon)$ is the retarded (advanced) uncoupled GF of the
left or right lead. The corresponding leaser (greater) self-energies
are given by
\begin{equation}
{\bf \Sigma}_{\mbox{\tiny
L,R}}^{\lessgtr}(\epsilon)=\left(\delta_{\lessgtr}-f(\epsilon,\mu_{\mbox{\tiny
\mbox{\tiny L,R}}}) \right) \left[{\bf \Sigma}_{\mbox{\tiny
L,R}}^r(\epsilon)-{\bf \Sigma}_{\mbox{\tiny L,R}}^a(\epsilon)\right],
\label{eq:SigmaLR}
\end{equation}
where $\delta_{\lessgtr}$ equals $0$ for $<$ and $1$ otherwise, and
$f(\epsilon,\mu)=\frac{1}{1+e^{\beta(\epsilon-\mu)}}$. The self-energy
arising from the interactions with the phonons is calculated using the
first Born approximation (FBA) and is given
by~\cite{Galperin2004,Mitra2004,Paulsson2005}:
\begin{equation}
\begin{split}
{\bf \Sigma}_{ph}^{r}(\epsilon) &= i\sum_{k} \int \frac{d\omega}{2\pi}
{\bf M}^{k} \left\{ D_{k}^{<}(\omega) {\bf g}^{r}(\epsilon-\omega) +
\right.\\ & \left. D_{k}^{r}(\omega){\bf g}^{<}(\epsilon-\omega) +
D_{k}^{r}(\omega) {\bf g}^{r}(\epsilon-\omega) \right\} {\bf M}^{k} ,
\end{split}
\label{eq: SEph}
\end{equation}
where the Hartree term has been omitted~\cite{Mitra2004}. The lesser
and greater self energies arising from the coupling to the phonons are
given by:
\begin{equation}
{\bf \Sigma}_{ph}^\lessgtr (\epsilon) = i\sum_{k} \int
\frac{d\omega}{2\pi} {\bf M}^{k} D_{k}^\lessgtr(\omega) {\bf
g}^\lessgtr(\epsilon-\omega) {\bf M}^{k}.
\label{eq:SEphLG}
\end{equation}
In the above equations, $D_{k}^{r,a}$ and $D_{k}^{\lessgtr}$ are the
uncoupled equilibrium retarded (advanced) and lesser (greater) GFs of
phonon mode $k$, respectively, ${\bf g}^{\lessgtr}(\epsilon)$ is the
lesser (greater) uncoupled electronic GF of the ring, and ${\bf
M}^{k}$ is the electron-phonon coupling matrix of mode $k$ (diagonal
in the $c_{j}$ basis).

\begin{figure}
\begin{center}
\includegraphics[width=7cm]{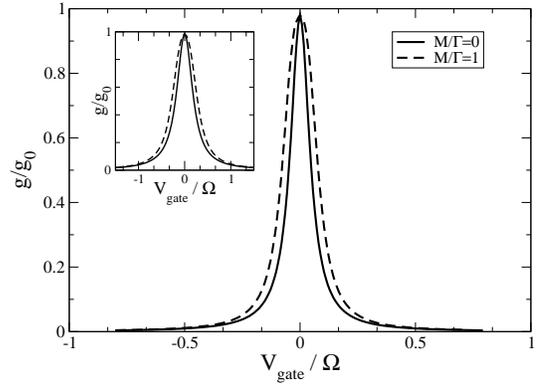}
\end{center}
\caption{Conduction as a function of the gate voltage at zero magnetic
flux with (dashed line) and without (solid line) electron-phonon
coupling.  Inset: A similar plot for the single resonant level model
described in Ref. \onlinecite{Mitra2004}.}
\label{fig:GVgate}
\end{figure}

We now turn to discuss the results of a specific realization of the
above model.  We consider a ring composed of $N=40$ sites.  The sites
are identical and contribute a single electron which is described by a
single Slater $s$-like orbital.  The coupling between the ring and the
leads is limited to the contact region.  For simplicity, the
electronic self-energies arising from this coupling are approximated
within the wide band limit. Specifically, we neglect the real-part of
the electronic self-energy and approximate ${\bf
\Gamma}_{L,R}(\epsilon)$ with matrices that are independent of energy,
where the only non-vanishing elements are the diagonal elements
($\Gamma_{L,R}$) corresponding to the two sites coupled to the left or
right lead.  The local phonon frequency $\Omega=0.0125$eV is
characteristic of a low frequency optical phonon in molecular devices.
Since our model does not include a secondary phonon bath required to
relax the energy from the optical phonons, we include a phonon energy
level broadening $\eta = 0.016\Omega$ which is included in the
uncoupled GFs of the phonon.  The coupling to each of the leads is
taken to be $\Gamma_{L}=\Gamma_{R}=4 \Omega$ such that the
magnetoconductance switching in the absence of electron-phonon
coupling is obtained at $\sim 5$~Tesla.

Before we address the effects of electron-phonon coupling on the
magnetoconductance properties of the system described above we will
analyze the role of a gate potential on the conductance.  In
Fig.~\ref{fig:GVgate} we plot the zero-bias conduction as a function
of a gate voltage with ($M/\Gamma=1$, where
$\Gamma=\Gamma_L+\Gamma_R$) and without ($M/\Gamma=0$) electron-phonon
coupling.  The gate voltage was modeled by an additional potential
$eV_{g} {\bf S}$, where ${\bf S}$ is the overlap matrix, that was
added to the ring hopping matrix element $t_{ij}$.  For comparison
(inset of Fig.~\ref{fig:GVgate}) we also include the results of a
single resonant level coupled to a single phonon with identical model
parameters used by Mitra {\em et al}.~\cite{Mitra2004}.

The two most significant observations are the expected broadening of
the conduction when the electron-phonon coupling is turned on and the
value of the zero-bias conduction ($g/g_{0}=1$, where $g_{0}=2e/h$ is
the quantum conductance) in the presence of electron-phonon coupling.
To better understand these results we rewrite the current for the case
that ${\bf \Gamma}_{L}(\epsilon) = {\bf \Gamma}_{R}(\epsilon) \equiv
{\bf \Gamma}(\epsilon)/2$ in the following way~\cite{Jauho1994}:
$I=\frac{2e}{h} \int d\epsilon \left(f(\epsilon - \mu_{L}) -
f(\epsilon - \mu_{R})\right) {\cal T}(\epsilon)$ where ${\cal
T}(\epsilon) = \frac{i}{4} \mbox{Tr} {\bf \Gamma}(\epsilon) \left({\bf
G}^{r}(\epsilon) - {\bf G}^{a}(\epsilon) \right)$.  Note that ${\cal
T}(\epsilon)$ is the transmission coefficient only when $M=0$.  In the
wide band limit, for the single resonant level model, ${\cal
T}(\epsilon)$ can be reduced to $\frac{\Gamma}{4} \frac{\Gamma -
2\Sigma_{ph,im}^{r}(\epsilon)} {\left(\epsilon - \epsilon_{0} -
\Sigma_{ph,re}^{r}(\epsilon) \right)^2 + \left(\Gamma/2 -
\Sigma_{ph,im}^{r}(\epsilon) \right)^2}$.  As a result of the fact
that $\Sigma_{ph,im}^{r}(0)=0$ at zero bias, the only inelastic
contribution to the conduction comes from the real part of the phonon
self-energy~\cite{Mitra2004}.  From this, it follows that even in the
presence of electron-phonon coupling, the maximal conduction is
$g_{\mbox{\tiny max}}/g_0=1$, as clearly can be seen in
Fig.~\ref{fig:GVgate} for both cases. It also immediately implies that
the main contribution to the broadening of the resonant conduction
peak comes from the real-part of the phonon self-energy, i.e., from
processes that lead to scattering phase shifts, but do not change the
lifetime of the state.

\begin{figure}
\begin{center}
\includegraphics[width=9cm]{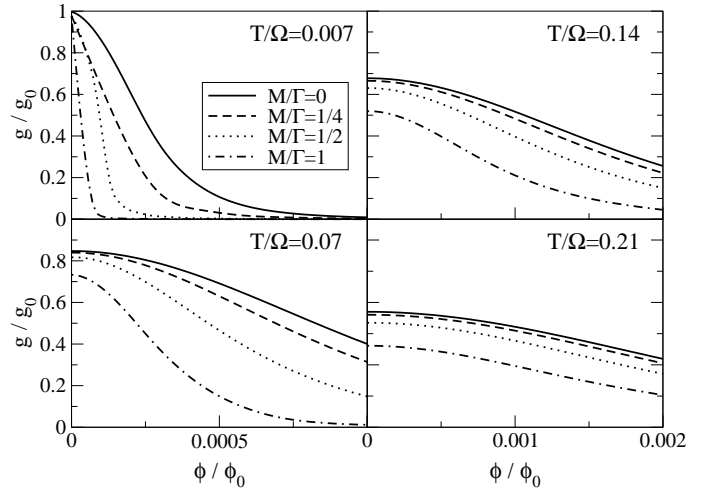}
\end{center}
\caption{Conductance as a function of magnetic flux for several values
of the electron-phonon coupling strength $M$ and for different
temperatures.}
\label{fig:gb}
\end{figure}

So far we have discussed the effect of electron-phonon coupling on the
zero-bias conduction as a function of a gate voltage. We now turn to
discuss the major result of the present study.  In Fig.~\ref{fig:gb}
we plot the magnetoconductance of the AB-ring for several values of
the electron-phonon couplings ($M$) and for different temperatures
($T$).  We focus on the low value of the magnetic flux $\phi=A B$,
where $A$ is the area of the ring and $B$ is taken perpendicular to
the ring plane.

The case $M=0$ for different systems was discussed in detail in our
previous studies, where the main goal was to establish the conditions
required to achieve negative magnetoconductance and magnetic switching
at low magnetic fields, despite the relatively large magnetic fields
required to complete a full AB period~\cite{Hod2004,Hod2005,Hod2006}.
The essential procedure described in [\onlinecite{Hod2004}] was to
weakly couple the AB-ring to the conducting leads and at the same time
to apply a gate potential to shift the position of the resonance state
such that conduction is maximized at $\phi/\phi_0=0$.  A manifestation
of these ideas is depicted in Fig.~\ref{fig:gb} for the case that
$M=0$ (solid curves), where the conduction is reduced from its maximal
value to a small value at a relatively low magnetic flux.  As
expected, we find that as the temperature is increased the maximal
value $g/g_0$ is decreased and the width of the magnetoconductance
peaks is increased linearly with $T$ for $M/\Gamma=0$ (with deviations
from linearity as $M/\Gamma$ is increased).  This increase in the width
with temperature is a result of resonant tunneling and the broadening
of the Fermi distributions as $T$ is varied.

\begin{figure}
\begin{center}
\includegraphics[width=7cm]{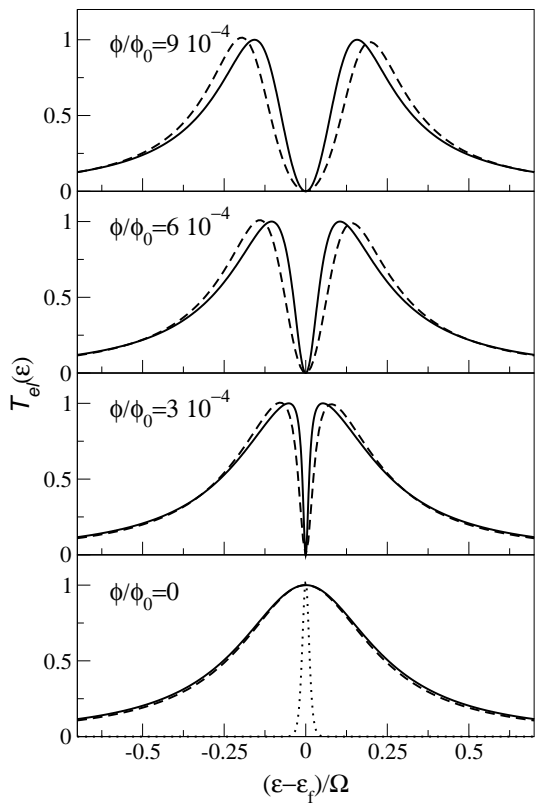}
\end{center}
\caption{Plots of ${\cal T}(\epsilon)$ as a function of energy for
$M/\Gamma=0$ (solid curves) and $M/\Gamma=\frac{1}{4}$ (dashed curves)
for different values of the magnetic flux. The dotted curve at the
lower panel shows $\frac{\partial}{\partial \mu}\Delta
f(\epsilon-\mu)$ at $T/\Omega=0.007$, where $\Delta f(\epsilon-\mu)$
is the difference in the Fermi distribution of the left and right
lead.}
\label{fig:ene}
\end{figure}

Turning to discuss the case of $M \ne 0$, one of the major questions
is related to the effects of electron-phonon coupling on the switching
capability of small AB-rings.  Based on the discussion of the results
shown in Fig.~\ref{fig:GVgate}, one might expect that an increase in
$M$ will lead to a broadening of the magnetoconductance peaks, thereby
increasing the value of the magnetic field required to switch a
nanometer AB-ring, and perhaps leads to unphysical values of $B$
required to reduced the conduction significantly.  As can be seen in
Fig.~\ref{fig:gb}, the numerical solution of the NEGF for $M \ne 0$
leads to a {\em reduction} of the width of the magnetoconductance
peaks, and the switching of the AB-ring is achieved at lower values of
the magnetic flux compared to the case where $M=0$.

This surprising observation can be explained in simple terms.  As
discussed above, even in the presence of electron-phonon coupling, the
maximal conduction at zero bias and zero temperature is
$g_{\mbox{\tiny max}}/g_0=1$, as clearly is the case for the results
shown in the upper left panel of Fig.~\ref{fig:gb} for
$\phi/\phi_0=0$.  For the symmetric ring of $N=4n$ the resonance
condition at $\phi/\phi_0=0$ is equivalent to the condition that
electrons entering the ring from left {\em interfere constructively}
when they exit the ring to the right~\cite{Hod2004,Hod2005}. This
picture also holds when $M \ne 0$, and the conduction takes a maximal
value at $\phi/\phi_0=0$. The application of a magnetic field leads to
destructive interference and increases the back scattering of
electrons.  This loss of phase is even more pronounced when inelastic
effects arising from electron-phonon coupling are included. In the
magnetoconductance this is translated to a more rapid loss of
conduction as a function of the magnetic field when $M$ is increased.

Mathematically, the rapid decay of the conduction with the magnetic
field as the electron-phonon coupling is increased can be explained by
analyzing the dependence of ${\cal T}(\epsilon)$.  In
Fig.~\ref{fig:ene} we plot ${\cal T}_{el}(\epsilon) =
\mbox{Tr}\left[{\bf \Gamma}_{\mbox \tiny L}(\epsilon){\bf
G}^{r}(\epsilon) {\bf \Gamma}_{\mbox \tiny R}(\epsilon){\bf
G}^{a}(\epsilon)\right]$, which is elastic (and dominant) contribution
to ${\cal T}(\epsilon)$, as a function of energy for several values of
$\phi/\phi_0$ for $M/\Gamma=0$ or $M/\Gamma=\frac{1}{4}$.  In the
lower panel we also plot the corresponding Fermi distribution window.
At $\phi/\phi_0=0$, ${\cal T}_{el}(\epsilon) \approx 1$ near the Fermi
energy ($\epsilon_f$), independent of $M$, and the conduction is $g /
g_{0} \approx 1$. The application of a small magnetic field results in
a split of ${\cal T}_{el}(\epsilon)$, where each peak corresponds to a
different circular state~\cite{Gefen2002}.  The separation between the
two peaks in the elastic limit $\Delta = (\epsilon_{2} - \epsilon_{1})
\propto \phi/\phi_0$ is proportional to the magnetic flux, where
$\epsilon_{1,2}$ are the corresponding energies of the two circular
states.  When inelastic terms are included, due to the fact that the
imaginary part of ${\bf \Sigma}_{ph}^{r}(\epsilon)$ is negligibly
small, the renormalized positions of the two peaks can be approximated
by $\epsilon^{*}_{1,2} = \epsilon_{1,2} +
\Sigma_{ph,re}^{r}(\epsilon_{1,2}) = \epsilon_{1,2} \pm
\Sigma_{ph,re}^{r}(\epsilon_{2})$, which implies that the renormalized
separation between the two peaks can be approximated by $\Delta^{*} =
(\epsilon^{*}_{2} - \epsilon^{*}_{1}) = \Delta + 2
\Sigma_{ph,re}^{r}(\epsilon_{2})$.  Therefore, as $M$ is increased
$\Delta^{*}$ is also increased, consistent with the numerical results
shown in Fig.~\ref{fig:ene}.

Similarly to the electrical gate, the magnetic field provides means to
externally control the conductance of a ring-shaped molecular
junction.  However, there are striking differences in the properties
of these two gauges.  This was illustrated previously in a
multi-terminal device, where the polarity of the magnetic field, which
couples to the electronic angular momentum, played a key role.  In the
present study we showed that there is also a fundamentally difference
with respect to inelastic effects.  While the conductance as a
function of the gate voltage broadens due to coupling to phonons it
actually narrows considerably in response to a magnetic field.  This
unexpected result was rationalized in terms of a rapid loss of the
phase of electrons at the exit channel arising from the coupling to
the phonons.  Mathematically, this effect was traced to the form of
the real part of the phonon self-energy that gives rise to scattering
phase shifts, but does not change the lifetime of the resonant level
through which conduction takes place.

We thank Abe Nitzan for many fruitful discussions. O.H. would like to
thank the generous financial support of the Rothschild and Fulbright
foundations. This work was supported by the Israel Science Foundation
(grants to E.R. and R.B.).  $^{\dagger}$Present address: Department of
Chemistry, Rice University, Houston, TX 77251-1892.

\end{document}